\documentclass[fleqn,usenatbib]{mnras}

\usepackage{newtxtext,newtxmath}

\usepackage[T1]{fontenc}

\DeclareRobustCommand{\VAN}[3]{#2}
\let\VANthebibliography\thebibliography
\def\thebibliography{\DeclareRobustCommand{\VAN}[3]{##3}\VANthebibliography}

\usepackage{amsmath}	
\usepackage{subcaption}
\usepackage[utf8]{inputenc}
\usepackage{multirow}
\usepackage{booktabs}
\usepackage{hyperref}
\usepackage{url}
\usepackage{datatool}
\usepackage{lineno}
\usepackage{ulem}
\usepackage{graphicx}	

\usepackage{xcolor}

\definecolor{forestgreen}{rgb}{0.1,0.49,0.07}
\definecolor{applegreen}{rgb}{0.55,0.71,0.0}
\definecolor{cadetblue}{rgb}{0.37, 0.62, 0.63}

\def\bl{Babcock--Leighton}

\newcommand{\Eq}[1]{Equation~(\ref{#1})}

\newcommand{\Sec}[1]{\S\ref{#1}}

\newcommand{\Fig}[1]{Figure~\ref{#1}}
\newcommand{\Tab}[1]{Table~\ref{#1}}

\DTLsetseparator{ = }

\usepackage{graphicx}
\graphicspath{{./figures/}}

\title[Is the Hemispheric Asymmetry of Monthly Sunspot Area an Irregular Process with Long-Term Memory?]{Is the Hemispheric Asymmetry of Monthly Sunspot Area an Irregular Process with Long-Term Memory?}

\author[Das, Ghosh\& Karak]{\parbox{\textwidth}{
        Ratul Das$^{1}$,
        Aparup Ghosh$^{2}$,
        Bidya Binay Karak$^{3,4}$\thanks{E-mail: karak.phy@iitbhu.ac.in}
        }
\vspace{0.4cm} \\
$^{1}$School of Physical Sciences, National Institute of Science Education and Research Bhubaneswar, Jatni, Khordha, Odisha, 752050, India\\
$^{2}$Department of Physical Sciences, Indian Institute of Science Education and Research Kolkata, Mohanpur, Nadia, West Bengal, 741246, India\\
$^{3}$Department of Physics, Indian Institute of Technology (Banaras Hindu University), Varanasi 221005, India\\
$^{4}$Max-Planck-Institut fu\"r Sonnensystemforschung, Justus-von-Liebig-Weg 3, D-37077 Go\"ttingen, Germany\\
}

\date{Accepted XXX. Received YYY; in original form ZZZ}

\pubyear{2021}

\begin{document}
\newcommand{\getval}[1]{\DTLfetch{data}{thekey}{#1}{thevalue}}
\newcommand{\geterr}[1]{\DTLfetch{data}{thekey}{#1}{errs}}

\label{firstpage}
\pagerange{\pageref{firstpage}--\pageref{lastpage}}
\maketitle

\begin{abstract}
The hemispheric asymmetry of the sunspot cycle is a real feature of the Sun.
However, its origin is still not well understood.
Here we perform nonlinear time series analysis of the sunspot area (and number) asymmetry to explore its dynamics. By measuring the correlation dimension of the sunspot area asymmetry, we conclude that there is no strange attractor in the data. Further computing Higuchi's dimension, we conclude that the hemispheric asymmetry is largely governed by stochastic noise.
However, the behaviour of Hurst exponent reveals that the time series is not completely determined by a memory-less stochastic noise, rather there is a long-term persistence, which can go beyond two solar cycles.  Therefore, our study suggests
that the hemispheric asymmetry of the sunspot cycle is predominantly originated
due to some irregular process in the solar dynamo. The long-term persistence in the solar cycle asymmetry suggests that the solar magnetic field has some memory in the convection zone.
\end{abstract}



\section{Introduction}
\label{sec:int}

The magnetic activity of Sun is not identical in two hemispheres---there is always
an asymmetry. This hemispheric asymmetry, also called the north-south asymmetry, has been
observed in the photospheric magnetic field \citep{MK04, McInt13, MK19} as well as in many proxies of the solar activity \citep{MKB17, GC09, norton14, Mordvinovetal20}.
The hemispheric asymmetry is a real feature of the solar cycle and is not an artefact of inaccurate or noisy observations \citep{carbo93}.

\citet{Bell61} found the evidence of hemispheric asymmetry in the number of major flares
and later \citet{Bell62} found a long-term asymmetry in the sunspot area data during Cycles 8--18.
\citet{Swinson86} found a peak in the northern hemispheric solar activity about two years after
sunspot minimum and a 22-years periodicity in the north-south asymmetry. \citet{Verma87} showed that the northern hemisphere is more active during Cycles 19 and 20. \citet{Li09} used group sunspot and sunspot
area data from 1996 to 2007 to show that the solar activity for cycle 23 is dominant in the
southern hemisphere; also see \citet{partha13} who extended this study to some part of cycle 24. A well-known sunspot asymmetry was observed during the Maunder minimum. Most of the sunspots were registered in the southern hemisphere \citep{SN94}.

It is believed that a hydromagnetic dynamo, operating in the solar convection zone, is responsible for the generation and maintenance of the large-scale magnetic field and the cycle of solar activity \citep{Pa55}.
In the current scenario of the solar dynamo \citep{Kar14a, Cha20, Hazra}, a toroidal component of the magnetic field
is largely generated due to the shearing of the poloidal component by the differential rotation. This toroidal field gives back to the poloidal one due to the decay and dispersal of tilted bipolar magnetic regions (BMRs)---
the so-called \bl\ process
and possibly due to the helical convection---the so-called $\alpha$ effect.
Meridional circulation and the small-scale convective flow play the role in transporting the magnetic field from the near-surface layer (the location of \bl\ process)
to the deeper convection zone, where the shearing process is efficient, and thus largely regulates the cycle period.

The turbulent nature of the helical convective flow---the main drive of the dynamo---is expected to make the magnetic field unequal in two hemispheres.
Thus, a hemispheric asymmetry
in the solar magnetic field is unavoidable. Furthermore, the tilts of BMRs, which primarily determines
the poloidal field, has a large scatter around Joy's law \citep{SK12, Wang15, Arlt16, Jha20}.
Thus the tilt scatter makes the poloidal magnetic field irregular and asymmetric \citep{LC17, KM17, Kar20}.
As the poloidal field is the seed for the toroidal field of the next solar cycle,
the asymmetry in the polar field is propagated in the solar cycle \citep{CCJ07}.
Dynamo models have shown that when the turbulent diffusion is sufficiently strong, the coupling between two hemispheres
tries to diminish the asymmetry introduced in the polar field and thus the asymmetry in the solar cycle may not persist for
several cycles \citep{CC06, GC09, Kar10, KM17}.
Dynamo models by including scatter in the BMR tilt \citep{LC17, KM17, KM18} or $\alpha$ term in the poloidal source \citep{OK13, KMB18, HN19} produce hemispheric asymmetry
in the magnetic cycle, which in some parameter regimes, are in agreement with observations. \citet{SC18} have shown that the random excitation of the quadrupole mode of dynamo by the stochastic fluctuations
in the \bl\ process can lead to an asymmetry in the observed magnetic field; also see \citet{Nepo19}.
Thus, all these previous results motivate us to explore whether the solar hemispheric asymmetry is governed by a low-dimensional chaotic process or stochastic process? Is there any long-term memory in the solar cycle asymmetry?

Nonlinear time series analysis is suitable to answer the above questions.
While there exist many such studies for the solar cycle data \citep[e.g.,][]{Ostriakov90, carbo94, jevti01, Letellier06,2009SoPh..260..441S}, only a few such studies are performed in the solar cycle asymmetry. 
\citet{carbo93} computed the correlation dimension of the asymmetry of daily sunspot area
during 1874--1989 and did not find any evidence of low-dimensional chaos. 
By computing the Higuchi's fractal dimension \citep{Higuchi88} and some nonlinear prediction method for the hemispheric asymmetry of sunspot number 
during 1947--1984, \citet{Watari96} concluded 
that the sunspot number asymmetry is highly irregular and not deterministic chaos.

In the present work, we shall utilize the maximum available sunspot area data (during 1874--2016) of hemispheric asymmetry and apply multiple nonlinear time series techniques to check the inherent nonlinear properties of the system. 
First, we shall compute the correlation dimension ($D_2$) to extract whether the data has any strange attractor in the data and thus this analysis will reveal the existence of any low-dimensional chaos in the underlying system \citep{GP83a}. 
Next, we shall compute a fractal dimension using the method given in \citet{Higuchi88} which provides a stable estimate of the fractal dimension when the data is more irregular and non-stationary. Higuchi's dimension will give another independent support of whether the asymmetry data is from a stochastic process or low dimensional chaos.  Finally, we shall 
compute the Hurst exponent \citep{1969WRR.....5..321M} to check whether the data has any persistent memory or not. 
The final conclusion will be presented in Section \Sec{sec:conclusion}.


\section{Observational data}
We use the monthly mean sunspot area data during May 1874 -- September 2016 obtained from the Royal Greenwich Observatory (RGO) \footnote{https://solarscience.msfc.nasa.gov/greenwch.shtml}. 
The RGO data have been the only available record of sunspot area separately in two hemispheres for the longer duration and this is routinely used in many studies of the solar activity \citep{Hat15}. 
RGO provides the monthly value of the average (over observed day) of the daily sunspot area in the unit of millionths of a hemisphere and it is evenly spaced.
We have also repeated our analyses with the newly available monthly mean hemispheric sunspot number data; \Sec{sec:HemSSN}.

\begin{figure}
\centering{\includegraphics[width=.99\columnwidth]{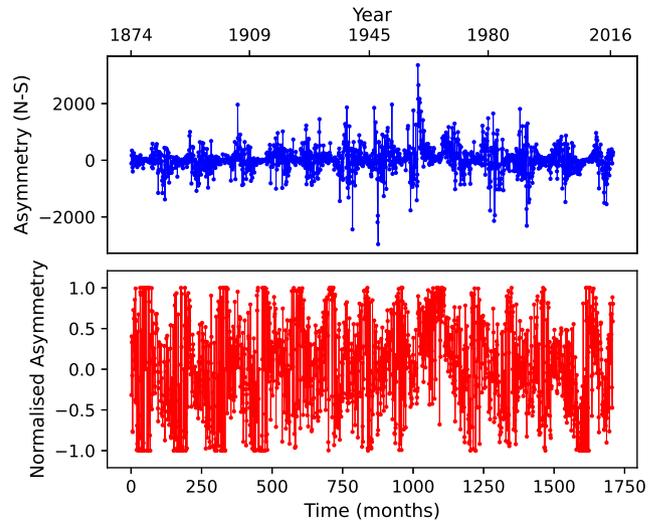}}
\caption{Time series of the hemispheric asymmetry of monthly average sunspot area 
(in unit of millionth of a solar hemisphere) 
as measured by $AS$ (top) and $AS_{Norm}$ (bottom).
}
\label{fig1}
\end{figure}

The easy way to measure the asymmetry is to take the difference in the values between two hemispheres \citep{Ballester2004, Chang2007}, 
i.e. asymmetry,
\begin{equation}
AS=A_N-A_S,
\label{eq:AS}
\end{equation}
where $A_N$ and $A_S$ are the monthly values of the sunspot area in the northern and southern hemispheres, respectively.
We note that during solar maxima, the difference becomes large in comparison to the value during minima and this causes a cyclic pattern in the asymmetry; see \Fig{fig1} top panel.
Furthermore, if a cycle is strong, then the asymmetry is large and vice versa. 
Therefore, in the literature \citep{carbo93, Oliver1996, Duchlev2001, GC09, partha13, Priy14}, 
the asymmetry is also measured by normalizing its strength, namely.
the normalized asymmetry,
\begin{equation}
AS_{Norm}=\frac{A_N-A_S}{A_N+A_S}.
\label{eq:normAS}
\end{equation}
When both $A_N$ 
and $A_S$ become equal to zero, we set $AS_{Norm}=0$.
We realized that this happens for 15 data points 
(which is less than $1\%$ of the total data). However, we 
discussed its effect in Section 4 by replacing these points 
with interpolated values.

We note that this $AS_{Norm}$ is a different time series 
than the $AS$; statistics of the data are different ( \Tab{table1}).
Again this definition of asymmetry is not satisfactory because during the solar minima, when the sunspot area becomes very small and this leads to increase in $AS_{Norm}$; see \Fig{fig1} bottom panel.
It was examined in \citet{Yi1992}, that dividing the difference of the hemispherical sunspot areas by the total sunspot area results in the appearance of a peak in the power spectrum 
between 11 and 12 years.

Due to such facts from the literature, where both definitions have been used to measure the solar cycle asymmetry, it becomes necessary to perform analyses of the time series of asymmetry using both the methods that we have discussed above.

\begin{table}
    \caption{
    Some statistics of the data. Total number of data points used: 1709. Symbols in second to third columns are as follows: $<S>$, the average of the time series; $rms$, the root mean-squared deviation, and $<{\rm PN}>=\sqrt{<S>}$.
    }
    \centering
    \begin{tabular}{lrrc}
    \hline
    \hline
    Data  & $<S>$   & $rms$~   & $<{\rm PN}>/rms$\\
    \hline
    North & 426.79 & 488.2 & 0.042 \\
    South & 409.43 & 470.1 & 0.043 \\ 
    AS    &  17.36 & 466.9 & 0.009 \\
    AS$_{\rm Norm}$ 
          &   0.01 &   0.6 & 0.225 \\ 
    \hline
    \end{tabular}
    \label{table1}
\end{table}

\section{Methods}
To identify the nonlinear properties of the hemispheric asymmetry of sunspot area time series, we shall compute three quantities for $AS$ and $AS_{Norm}$, namely, the Correlation dimension, Higuchi's dimension, and Hurst exponent. Below, we discuss briefly how to compute these quantities.

\subsection{Correlation dimension}
First, we shall apply a method of time series analysis to distinguish the random noise in the underlying system from the low dimensional chaos.
This method is to measure a fractal dimension of the strange attractor
in the system which is popularly known as the correlation dimension ($D_2$).
The method for obtaining $D_2$ has been given in \citet{GP83a,GP83b}, also see \citet{takens81,packard80}.
This has also been used in many astrophysical applications \citep[e.g.,][]{schreiber99,misra06,KDM10}, 
including identifying chaotic dynamics of the solar cycle data \citep{OU90,carbo94,jevti01}.
In this method, we construct a $M$ dimensional phase space using our time series $X(i)$ (where $i= 1, 2, 3 .....N$); $X(i)$ is $AS$ or $AS_{\rm norm}$ in our case. 
In this space, any vector has the following form:
\begin{equation}
x_{i} = [X(i), X(i+\tau), ........., X(i+(M-1)\tau)].
\label{eq:vec}
\end{equation}
The time delay, $\tau$ is chosen in such a way that each component becomes independent of each other. 
We find that at $\tau = 3$~months, the auto-correlation of the data falls below $1/e$ and thus we
set this value for $\tau$ in our analysis \citep[see][for a detailed discussion on choosing $\tau$]{carbo94}. We checked that our results do not change abruptly
if we increase $\tau$. 

Next, the correlation function $C_M(r)$ is given as
\begin{equation} 
C_M(r)=\frac{1}{N(N_c-1)}\sum_{i=1}^{N} \sum_{(j,j\neq i)}^{N_c}\Theta(r-|x_{i}-x_{j}|),
\end{equation} 
where $x_{i}$ is a reconstructed vector; \Eq{eq:vec}, $\Theta$ is Heaviside function 
($\Theta(z)$ = 1 if $z\geq 0$ and 0 if $z < 0$), 
$N$ the total number of points and $N_c$ is the number 
of centers.
Essentially, $C_M(r)$ gives the number of points that are within a distance $r$ from the centre, 
averaged over all the centres.
Then for small $r$, ${D_{2}(M)}$ is given by 
\begin{equation}
{D_{2}(M)}=  \frac{\rm d log(C_M(r))}{\rm d log(r)}.
\label{eq:slope}
\end{equation}

To compute $C_M(r)$, we divide the entire phase space into $M$ cubes of length
$r$ around a point and count the average number of  points. To avoid
the edge effects due to the finite number of data points, we compute 
$C_M(r)$ in the range $ r_{min}<r<r_{max}$. Here $r_{min}$ is chosen when
$C_M(r)$ is just greater than one and $r_{max}$ is taken in such a way that all $M$
cubes remain within the embedding space.
For a fixed value of $M$,
$D_{2}$ 
is computed for different values of $r$ in the linear region of ${\rm log}( C_M( r))$ versus ${\rm log}( r)$ plot using
\Eq{eq:slope}.
The average of all these values will give our final $D_{2}$
and the mean standard deviation over the average value gives the error on $D_{2}$.
The whole calculation is repeated for different values of $M$.

The value of $D_{2}$ should increase initially with
the increase of $M$. However, if the time series is obtained from a low dimensional chaotic system, then
$D_{2}$ tends to saturate above a certain value of $M$. In contrast, for a stochastic system, $D_{2}$ keeps on increasing
with $M$. That is the number of dimensions needed to describe the system in the phase space is infinitely large in a stochastic system. In that case, 
$D_{2} \approx M$, for all $M$. Thus, the variation of $D_2$ with $M$ is used to distinguish between the random noise vs low dimensional chaos.

\subsection{Higuchi's dimension}
Previously, many methods for finding stable estimations of the power-law spectral index have been discussed in the literature. Here, we follow the method given in \citet{Higuchi88} to calculate the fractal dimension $D$ of the asymmetry series. This method is helpful in providing a stable estimate of the fractal dimension of the asymmetry time series.

We recall that $X(i)$ is our time series ($i =1, 2, 3,..., N$; $N$ is the total number of observations taken at a regular interval) and thus,
\begin{equation} 
X(i): X(1), X(2), X(3), ..., X(N)
\end{equation}
From this, a new time series is constructed in the following manner: 
\begin{equation} 
X^m_\tau: X(m), X(m+\tau), X(m+2\tau), ..., X\left(m+\left[\frac{N-m}{\tau}\right]\tau\right)
\end{equation}
where $m=1,2,...,\tau$, and $[\:]$ denotes Gauss's notation. The length $L_{m}(\tau)$ of the curve associated to each $X^m_\tau$ is defined as:
\begin{equation} 
L_{m}(\tau)=\left\{\left(\sum_{i=1}^{\left[\frac{N-m}{\tau}\right]}|X(m+i\tau)-X(m+(i-1)\tau)|\right)\frac{N-1}{\left[\frac{N-m}{\tau}\right]\tau}\right\}\frac{1}{\tau},
\end{equation}
The average value of the time series length $\langle{L(\tau)}\rangle$ for a given value of $\tau$ is defined as the average of $\tau$ sets of $L_{m}(\tau)$. If $\langle{L(\tau)}\rangle \propto \tau^{-D}$, for the range $\tau_{min} < \tau < \tau_{max}$ then the time series is a fractal and has a dimension $D$ for that range of $\tau$. We find $\langle{L(\tau)}\rangle$ for $\tau=2$ to $\tau=55$ for our analysis, for both $AS$ and for $AS_{Norm}$. 

\subsection{Hurst exponent}
Here, we attempt to find the Hurst exponent, $H$, which characterises the persistence of a times series to examine whether the non-periodic variations in the asymmetry time series are a result of a white noise process, an anti-correlated random process, or a correlated random process. We borrow the discussion regarding the Hurst exponent in the context of solar activity data from \citet{1994SoPh..149..395R} and \citet{2009SoPh..260..441S}. We use the $R/S$ method as given in \citet{1969WRR.....5..321M} to obtain the Hurst exponent.

We choose a temporal window $\tau$, where $\tau_{t}>\tau>N$, and $\tau_{t}$ is the Theiler window \citep{1986PhRvA..34.2427T}, to make subsets of time series $X(i)$ as follows:
\begin{equation} 
x_i(\tau); X(t_0), X(t_0+1), X(t_0+2), ..., X(t_0+\tau-1),
\end{equation}
where $t_0 = 1, 2, ..., N-\tau+1$. It is important to note that the choice of window used in finding the $R/S$ values for each given $\tau$ is rather important, since non-overlapping windows produce $R/S$ values from comparatively small sample sizes. Lesser number of windows could possibly provide inaccurate values for $H$. 
Now we denote the average of these subsets as:
\begin{equation} 
\bar{x}(t_0,\tau)=\frac{1}{\tau}\sum_{i=t_0}^{t_0+\tau-1}x_i(\tau).
\end{equation}
Let, $S(t_0,\tau)$ be the standard deviation of $x_i(\tau)$ for the window $\tau$ as follows:
\begin{equation} 
S(t_0,\tau)=\sqrt{\frac{1}{\tau-1}\sum_{i=t_0}^{t_0+\tau-1}\left[x^{t_0}_\tau(i)-\bar{x}(t_0,\tau)\right]^2}.
\end{equation}
Now, we define a set of new variables $y_i(t_0,\tau)$, which is the set of cumulative deviations from the mean of $x_i(\tau)$
\begin{equation} 
y_i(t_0,\tau)=\sum_{k=t_0}^{t_0+i-1}\left[x^{t_0}_\tau(k)-\bar{x}(t_0,\tau)\right],
\end{equation}
and hence, the range $R$ of $y_i(t_0,\tau)$ is obtained as:
\begin{equation} 
R(t_0,\tau)=\max_{1\leq i\leq \tau}y_i(t_0,\tau) - \min_{1\leq i\leq \tau}y_i(t_0,\tau).
\end{equation}
This allows us to define the rescaled range measure $R/S$ as:
\begin{equation} 
(R/S)(t_0,\tau)=\frac{R(t_0,\tau)}{S(t_0,\tau)}.
\end{equation}
Calculating the $R/S$ values for each temporal window by moving from $t_0=1$ to $t_0=N-\tau+1$ for window size $\tau$, the rescaled range for $\tau$ is then given as the average of these values
\begin{equation} 
(R/S)=\frac{1}{N-\tau+1}\sum_{t_0}(R/S)(t_0,\tau).
\end{equation}
It was observed that the rescaled range for a time window is proportional to $\tau^H$
\begin{equation}
(R/S)_\tau=k\tau^H
\end{equation}
where $k$ is the proportionality constant, and $H$ is the Hurst exponent. To obtain the value of the Hurst exponent, $R/S$ values are plotted for $\tau=11$ to $\tau=1709$ for the Hurst Exponent analysis, for both $AS$ and $AS_{Norm}$.

\begin{figure}
\centering
\includegraphics[width=1.0\columnwidth]{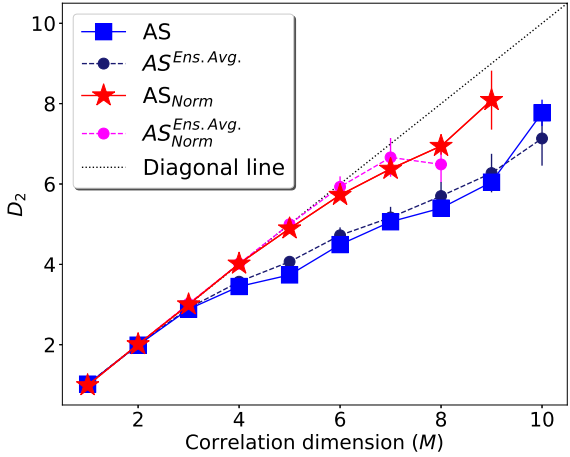}
\caption{
Variation of correlation dimension ($D_2$) with the embedding dimension ($M$) for $AS$ (square points) and $AS_{\rm Norm}$ (asterisks). The dotted line
along the diagonal of the figure indicates the $D2$ variation as expected from an ideal stochastic process. $AS^{Ens.Avg.}$ and $AS_{Norm}^{Ens.Avg.}$ (round points) shows the variation averaged over an ensemble of 1000 AS \& $AS_{Norm}$ time series, respectively (see Section \ref{Sec_Error_estimates}).
}
\label{fig:MvsD2}
\end{figure}

\begin{figure}
\centering
\includegraphics[width=.95\columnwidth]{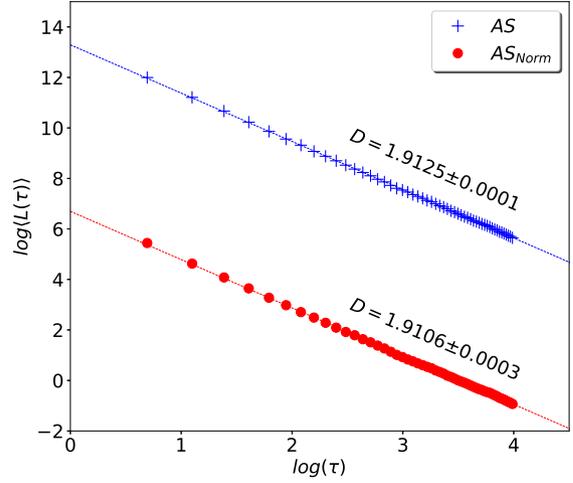}
\caption{
Variation of the length ($\langle{L(\tau)}\rangle$) with the time interval ($\tau$) for $AS$ and $AS_{Norm}$.
}
\label{fig:higuchi}
\end{figure}

\begin{figure}
\centering
\includegraphics[width=.95\columnwidth]{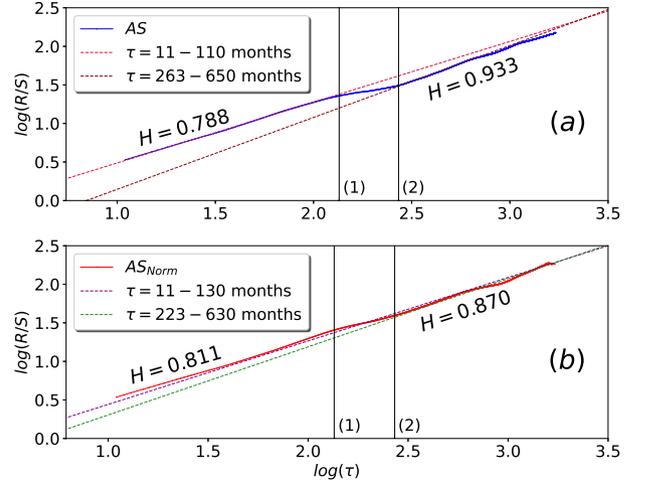}
\caption{
Variation of $R/S$ with time interval ($\tau$) for $AS$ (top) and $AS_{Norm}$ (bottom). $\tau=135$ months and $\tau=271$ months are indicated as (1) and (2).
}
\label{fig:HurstExp}
\end{figure}

A white noise process or a random walk process is defined by a Hurst exponent of $H=0.5$. When the time series has $H>0.5$, it is said to be persistent. Persistence is defined as the tendency for the process to have a memory of the previous step. That is, if there was an increase in the value of the time series, the following step would be more likely to have an increase as well. In such a case, the time series would cover more ``distance'' than a random walk would. Whereas, if $0<H<0.5$, then the time series is said to be anti-persistent. That is, an increase in the value of the time series is more likely to be followed by a decrease, and vice-versa. And in opposition to a persistent case, the time series would cover less ``distance'' than a random walk. 

There are other methods to determine the Hurst exponent, and the value that is obtained is sensitive to the method used \citep{weron02}. In order to provide a  confidence in the estimate of the Hurst exponent, and to ensure that the result is not method dependent, we shall compute Hurst exponent in two more methods, namely,  Detrended Fluctuation Analysis and Periodogram Regression. As these methods are well described in 
the literature, we shall describe them briefly in Appendix~A.


\section{Results}
The variation of $D_2$ as function of $M$ for the sunspot area asymmetry is shown in \Fig{fig:MvsD2}. We find that the asymmetry and normalized asymmetry do not show the same behaviour.   
Nevertheless, in both cases, $D_2$  increases with an increase of $M$. The lack of saturation in $D_2$ implies that the sunspot area asymmetry is not governed by low-dimensional chaos,
rather it might be driven by a high-dimensional or stochastic process. 
This conclusion is in general agreement with \citet{carbo93} who also did not find the evidence of low-dimensional chaos in the asymmetry of sunspot area
data during 1874--1989.

To confirm that the solar cycle asymmetry is really governed by stochastic
or high-dimensional chaos, we observe the value of Higuchi's dimension ($D$). 
As seen from \Fig{fig:higuchi}, 
 for $AS$, $D = 1.9125 \pm 0.0001$, and for $AS_{Norm}$, $D = 1.9106 \pm 0.0003$. 
We know that when the value of $D$ for a curve is close to $2$, the curve behaves nearly like a surface, i.e., the curve is close to a space-filling curve. Hence, the self-similar nature for time series, i.e., the hallmark of low-dimensional chaos is absent. Therefore, we conclude that the process that generates the hemispheric asymmetry of sunspot area is very likely to be the result of an irregular or stochastic process.

Finally, we explore the memory of these irregular asymmetry data by computing 
the Hurst exponent ($H$). 
In \Fig{fig:HurstExp}, we show the log-log plots for $R/S$ against $\tau$.
The slope of this curve gives the H value. We, however, see two distinguishable linear scaling regimes, and hence one value of H for all $\tau$ is not adequately representing the data. The previous study for sunspot number cycle also indicted two distinct
regimes \citep{2009SoPh..260..441S}.

For $AS$, in the range: $\tau=11$--110 months, we find $H=0.79$, 
while for $\tau=263$--650 months, we obtain $H=0.94$ 
(\Tab{table2}).
In between these two regimes, there is a small region during 135--271 months (marked by vertical lines in  (\Fig{fig:HurstExp}a)) with a weaker slope, which is possibly linked to a period of lower persistence in the trend.
For $AS_{Norm}$ (\Fig{fig:HurstExp}b), in the range $\tau=11$--130 months, we find $H=0.81$, while in the range: $\tau=233$--630 months, we get $H=0.87$. In this case, the change in the slope happens very slowly. 


\begin{figure}
\centering
\includegraphics[width=.95\columnwidth]{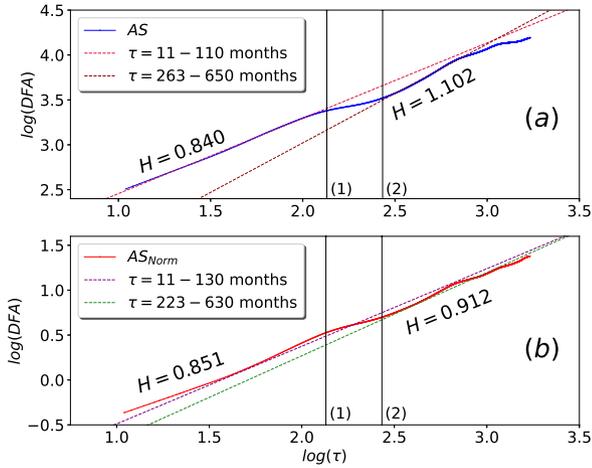}
\caption{ Variation of $DFA$ with time interval ($\tau$) for $AS$ (top) and $AS_{Norm}$ (bottom). $\tau = 135$ and $\tau = 271$ months are indicated as (1) and (2).
}
\label{fig:HurstExp_DFA}
\end{figure}

\begin{figure}
\centering
\includegraphics[width=.95\columnwidth]{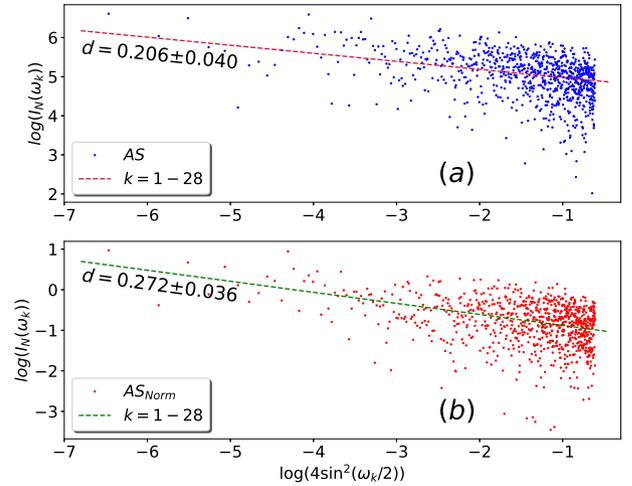}
\caption{
Periodogram Geweke-Porter-Hudak (GPH) Method for $AS$ (top) and $AS_{Norm}$ (bottom).
}
\label{fig:HurstExp_PR}
\end{figure}

In the Detrended Fluctuation Analysis (DFA), we get similar results but 
the values of $H$ are slightly larger (Figure \ref{fig:HurstExp_DFA}). 
For $AS$, in the range: $\tau = 1$--110 months, we find $H = 0.84$, while for $\tau = 263-650$ months, we obtain $H = 1.10$. 
Now, as discussed in \cite{BS12, CL17}, we can point out that a value of $H$ above 1 is not impossible. It is a consequence of non-stationarity or a trend not being fully removed from the data. The reliability of DFA as a valid method has been questioned on similar grounds before. But since our R/S analysis still backs up the general result that the latter regime has a higher slope, we can safely negate the effect that this inconsistency may cause.
(\Tab{table2}). 

For $AS_{Norm}$, in the range: $\tau = 11-130$ months, we find $H = 0.85$, while in the range: $\tau = 223 - 630$ months, we get $H = 0.91$.


Using the Method of Periodogram Regression (Figure \ref{fig:HurstExp_PR}), for $AS$, we obtain $H (= 0.5+0.21) = 0.71$, while for 
$AS_{Norm}$, we obtain $H (= 0.5+0.27) = 0.77$.
Unlike the case in $R/S$ Method and Detrended Fluctuation Analysis, the periodogram (Figure \ref{fig:HurstExp_PR}) does not show two distinct linear scaling regimes. However, the Hurst exponent being larger than 0.5, we can safely conclude that $AS$ and $AS_{Norm}$ time series are persistent in nature and the degree of persistence increases with the time-scale. Larger value of $H$ for 
$\tau \gtrapprox $ 22 years ($=264$ months), suggest that
the memory of solar cycle asymmetry persists at least for two cycles.

We have obtained $H$ 
and $D$ 
graphically, using the Least Squares method for all of our plots; see e.g., \Fig{fig:higuchi}. 
 We have also applied Bayesian linear regression to all our power law fits \citep{2004ApJ...609.1134W} using the justification towards a Bayesian method of fitting as opposed to a frequentist method, as explained in \citet{D'Huys2016}. We use the Python module, PyMC3 \citep{Salvatier2016} for this. 
 We find that the differences in results lie in the range of $10^{-4}$ -- $10^{-3}$, except in the case of of the periodogram method. In that case, for $AS$ and $AS_{Norm}$ we obtain the value of the Hurst exponents as $0.707\pm0.179$ and $0.761\pm0.132$, respectively using Bayesian linear regression, while these values obtained from previous Least Squares method are $0.706$ and $0.772$, respectively. 

We recall that while computing $AS_{\rm Norm}$ using \Eq{eq:normAS},
we took $AS_{\rm Norm} = 0$ when both $A_N$ and $A_S$ are zero. Instead of this,
if we replace these points by interpolating the neighbouring points, then 
this so-called zero-replacement strategy affects our computed results only marginally. (For example, from this zero-replacement $AS_{\rm Norm}$ data, 
the computed value of $D$ is 1.9077 and the values of
$H$ are 0.8110 \& 0.8693 (R/S method), 0.8469 \& 0.9074 (DFA), and 0.7718 (PR). Compare these values with
the corresponding values in \Tab{table2}.)

\begin{table*}
\centering
\caption{
Summary of results obtained for Higuchi's dimension and Hurst exponents. 
In the forth and sixth columns, the values are the means and the standard deviations (errors) of the results obtained from an ensemble of 1000 AS \& AS$_{Norm}$ time series, respectively (see text).}

\begin{tabular}{llcccc}
\hline \hline
\multicolumn{1}{l}{Method} & \multicolumn{1}{l}{Window} & \multicolumn{4}{c}{Time Series}                                                                        \\ \cline{3-6} 
                           & (months)                                                                      & AS                   & Ensemble of AS & AS$_{Norm}$ &  Ensemble of AS$_{Norm}$          \\ \hline
        Higuchi                   &          2-55                                                                     & \multicolumn{1}{l}{1.912} & \multicolumn{1}{c}{1.915$\pm$0.002}                                     & 1.911 & \multicolumn{1}{c}{1.916$\pm$0.002} \\ \hline
\multicolumn{1}{c}{Hurst (R$/$S)}     &11-110                                                    &    0.788                  &           0.783$\pm$0.008                                               &   0.811  & \multicolumn{1}{c}{0.813$\pm$0.007} \\
                           & 263-650                                                                       & \multicolumn{1}{l}{0.933} & \multicolumn{1}{c}{0.929$\pm$0.019}                                     &0.871 & \multicolumn{1}{c}{0.879$\pm$0.017} \\ \hline
\multicolumn{1}{c}{Hurst (DFA)}    & 11-110                                                    &         0.840              &        0.834$\pm$0.010                                                  &       0.851    & 0.851$\pm$0.008          \\
                           & 263-650                                                                       & \multicolumn{1}{l}{1.102} & \multicolumn{1}{c}{1.096$\pm$0.022}                                     & 0.912 & \multicolumn{1}{c}{0.920$\pm$0.018} \\ \hline
\multicolumn{1}{c}{Hurst (PR)}     & K = 28                                                    &    0.706                  &     0.704$\pm$0.039                                                     &      0.772   & 0.775$\pm$0.028             \\ \hline
\end{tabular}

\label{table2}
\end{table*}

\section{Error Estimates}
\label{Sec_Error_estimates}
In our study, we used the monthly averaged hemispheric sunspot area as recorded in RGO. Unfortunately, in these data, no error information is given. Therefore we cannot make a direct estimate of error in our computed results. However, using the daily sunspot area 
data$^1$,
we can make some estimate of the errors in the following way 
using a Bootstrapping technique \citep{bootstrap}.
Let us consider the daily sunspot area data of one month for both the northern and southern hemispheres. We produce 100 resampled datasets with the same size as the number of days for this month by randomly selecting daily pairs of values for north and south, and then computing the corresponding $AS$ and $AS_{Norm}$ for the resamples. We compute the mean for all of these resamples. And then, we compute {\it  the mean} ($\mu$) and {\it the standard deviation} ($\sigma$) of the means of all the resampled datasets for a month. 
It can be easily seen that this mean is not necessarily the same as what we have used in our earlier analyses.
With these $\mu$ and $\sigma$, we produce an ensemble of 1000 data points (deviate) from a Gaussian distribution and repeat this for all the months to get 1000 time series. Finally, perform our all the nonlinear time series analyses with this ensemble. 

Black and magenta filled circles connecting dashed lines in \Fig{fig:MvsD2} show the average $D_2$ behaviour of the ensemble of 1000 $AS$ and $AS_{Norm}$ time series. The error bar represents the $\sigma$ of the computed $D_2$ of 1000 time series. In \Tab{table2}, forth and sixth columns show the mean and error (standard deviation) of $D$ and $H$ from the ensembles.

We clearly see that the mean values of the computed quantities ($D_2$, $D$, and $H$)  of the ensemble of 1000 $AS$ 
\& $AS_{Norm}$
time series are not too far from the ones computed from the original monthly mean time series. 
The values of $\sigma$ of the ensemble are also reasonably low, with the quantities being less than 1 standard deviation away from the computed values in most cases.





\section{Discussion and Conclusion}
\label{sec:conclusion}
We have explored some nonlinear properties of the underlying process behind the solar cycle asymmetry using nonlinear time series analysis. For this, we have used the hemispheric monthly sunspot area and number time series, which are the best proxies of the Sun's large-scale magnetic flux available for a longer duration.

Following the literature, 
 solar cycle asymmetry has been measured in two ways, namely, the hemispheric difference $AS$ and the normalized hemispheric asymmetry $AS_{Norm}$.
We have used three methods of time series analyses to characterise the data.

From the analysis of the correlation dimension $D_2$, we find that the value of $D_2$ does not saturate for higher values of $M$. This indicates that there is no underlying presence of a low-dimensional chaotic attractor that could govern the asymmetry of sunspot area data, in agreement with the conclusion obtained in \citet{carbo93}. 
In other words, we can expect that the 
asymmetry is likely to be 
produced by irregular process.

In our fractal analysis, we see that the value obtained for the Higuchi's fractal dimension ($D$) is close to $2$, which implies that a stochastic process or possibly a high-dimensional chaotic process is the cause of the asymmetry.

In all three methods of computation of Hurst exponent, 
we 
find the value of Hurst exponent $H$ 
is above 0.7 for $AS$ data and a little larger 
for $AS_{\rm Norm}$. 
We find multiple values of $H$ for the same time series. Its value
decreases slightly after about 11 years (one cycle period) and then increases for windows larger than about 22 years. This change in the value of H and thus the persistence is more prominent in $AS$. In general, a memory can be observed for the asymmetry time series and it is larger in long-time scale (beyond 22 years; two cycles).
From our analysis, we conclude that the monthly hemispheric asymmetry of sunspot area is dictated by a stochastic process with some amount of long-memory.


The results from hemispheric sunspot number, which is recently made available
by \citet{Veronig21} during 1874--2020, also shows similar behaviour (during the period May 1874 - September 2016) as 
that found in the hemispheric area data; \Sec{sec:HemSSN}.

Stochastic nature of hemispheric asymmetry supports the previous theoretical studies \citep[e.g.,][]{GC09, OK13, KM17, SC18, HN19, Nepo19} which explains the solar cycle asymmetry to be caused by the
irregularity involved in the helical convective flow, and in particular the randomness involved in the Babcock--Leighton process (e.g., 
in the form of tilt of BMRs, emergence rate, meridional flow). Further, the presence of some long-term memory in asymmetry
time series supports the existence of a finite memory of the sun's magnetic field, which is possibly determined by the turbulent diffusion and pumping \citep{CC06, KN12, KM17, KK21}.

\section*{Acknowledgement}
B.B.K. thanks Banibrata Mukhopadhyay, Jayanta Dutta and Vinita Suyal for many discussion on time series analysis and help in writing codes during his PhD time. 
He also thanks Bibhuti Kumar Jha and Prasun Dutta for the discussion on error analyses.
Authors thank the anonymous referee who provided us valuable feedback on the earlier versions of this paper.
B.B.K. acknowledges the funding from Department of Science and Technology (SERB/DST), India through the Ramanujan Fellowship (project no SB/S2/RJN-017/2018) 
and the support provided by the Alexander von Humboldt Foundation during a part of this project. A.G. acknowledges Kishore Vaigyanik Protsahan Yojana (KVPY) for scholarship.
R.D. acknowledges the DAE Incentive Scheme for Holistic Science Education and Augmentation (DISHA) for scholarship.

\section*{DATA AVAILABILITY}
Sunspot area data used in the present study is obtained from the Royal Greenwich Observatory; http://solarscience.msfc.nasa.gov/greenwch.shtml.
Data of our analyses presented in the article will be shared upon reasonable request to the corresponding author.


\bibliographystyle{mnras}
\bibliography{paper} 



\appendix
\section{Hurst exponent using different methods}

In this section, we will determine the Hurst exponent by two other methods, namely, Detrended Fluctuation Analysis and Periodogram Regression.

The method of \textit{Detrended Fluctuation Analysis} proposed by \citet{peng94} and also discussed in \cite{weron02} can be summarized as follows. We choose a temporal window $\tau$, where $\tau_0 \le \tau \le N$. Here, $\tau_0$ plays a similar role as the Theiler window plays in the $R/S$ method. For this $\tau$, we can divide the time series $X(t)$ into $d(= N-\tau+1)$ subseries
\begin{equation}
x_i (\tau): X(i), X(i+1), ... , X(i+\tau-1)
\end{equation}
for $i = 1, 2, 3, ... , d$. Note that each subseries $x_i (\tau)$ is of length $\tau$. Now, for each subseries $x_i (\tau)$, we create a cumulative time series
\begin{equation}
Y_i (\tau) : X(i), \sum_{j = 0}^{1} X(i+j), ... , \sum_{j = 0}^{\tau-1} X(i+j).
\end{equation}
Now, we fit a least-squares line $\overline{Y}_i (\tau, x) = a_i x + b_i$ where, $x = 1, 2, ... , \tau$, to $\{Y_i (\tau)\}$. Let, $S_i (\tau)$ be the root mean square fluctuation (i.e. standard deviation) of the Integrated and Detrended time series, given by
\begin{equation}
S_i (\tau) = \sqrt{\frac{1}{\tau} \sum_{x = 1}^{\tau} (Y_{i, x} (\tau) - a_i x - b_i)^2 }
\end{equation}
where, $Y_{i, x} (\tau)$ represents the x-th element of the cumulative time series $Y_i (\tau)$. Finally, we calculate the mean value of the root mean square fluctuation for all subseries of length $\tau$, $(DFA)_{\tau}$, given by
\begin{equation}
(DFA)_{\tau} = \frac{1}{d} \sum_{i = 1}^{d} S_i (\tau)
\end{equation}
Similar to the case of the $R/S$ analysis, a linear relationship on a double logarithmic paper of $(DFA)_{\tau}$ against $\tau$ indicates 
the presence of a power-law scaling
\begin{equation}
(DFA)_{\tau} = k \tau^H
\end{equation}
where, $k$ is the proportionality constant and $H$ is the Hurst exponent. To obtain the values of the Hurst exponent, $(DFA)$ values are plotted for $\tau = 11$--$1709$ months for both $AS$ and $AS_{Norm}$. \\

Now, we will discuss the method of \textit{Periodogram Regression} proposed by \citet{geweke83}. It is calcualted from 
the slope of the spectral density function of a fractionally integrated series at low frequencies. 

For the time series, we start out by calculating the periodogram, which is a sample analogue of the spectral density. Given the time series $X(i)$ of length $N$, the periodogram is obtained as 
\begin{equation}
I_N (\omega_k) = \frac{1}{N} \left\lvert \sum_{t = 1}^N X(t) e^{-i 2 \pi (t-1) \omega_k } \right\rvert ^2
\end{equation}
where, $\omega_k = k/N, k = 1, 2, ... , [N/2]$ and 
$[X]$
denotes the greatest integer less than or equal to 
$X$
Note that, $I_L$ is the squared absolute value of the Fourier transform. The next step is to run a linear regression 
\begin{equation}
\log \{ I_N (\omega_k) \} = a - \hat{d} \log \{ 4 \sin^2 (\omega_k/2) \} + \epsilon_k
\end{equation}
at low Fourier frequencies 
$\omega_k, k = 1, 2, ... , K \le [L/2]$. 
$\omega_k, k = 1, 2, ... , K \le [N/2]$. 
The least squares estimate of the slope yields the differencing parameter $d = \hat{d}$ and $H = \hat{d} + 0.5$. 
Next is how to set the value of $K$.
Since, the differencing parameter $d$ is
sensitive to the choice of $K$, we decided to keep it smaller than the standard value of $N^{0.5}$. We use $K = [N^{0.45}] = 28$ in our analysis to determine the Hurst exponent, keeping in mind smaller powers of $N$ introduce large estimation errors. Once the differencing parameter $d$ was obtained, the Hurst exponent was obtained as $H = d + 0.5$. 

\section{Results from Hemispheric Sunspot Numbers}
\label{sec:HemSSN}
\begin{figure}
\centering
\includegraphics[width=1.1\columnwidth]{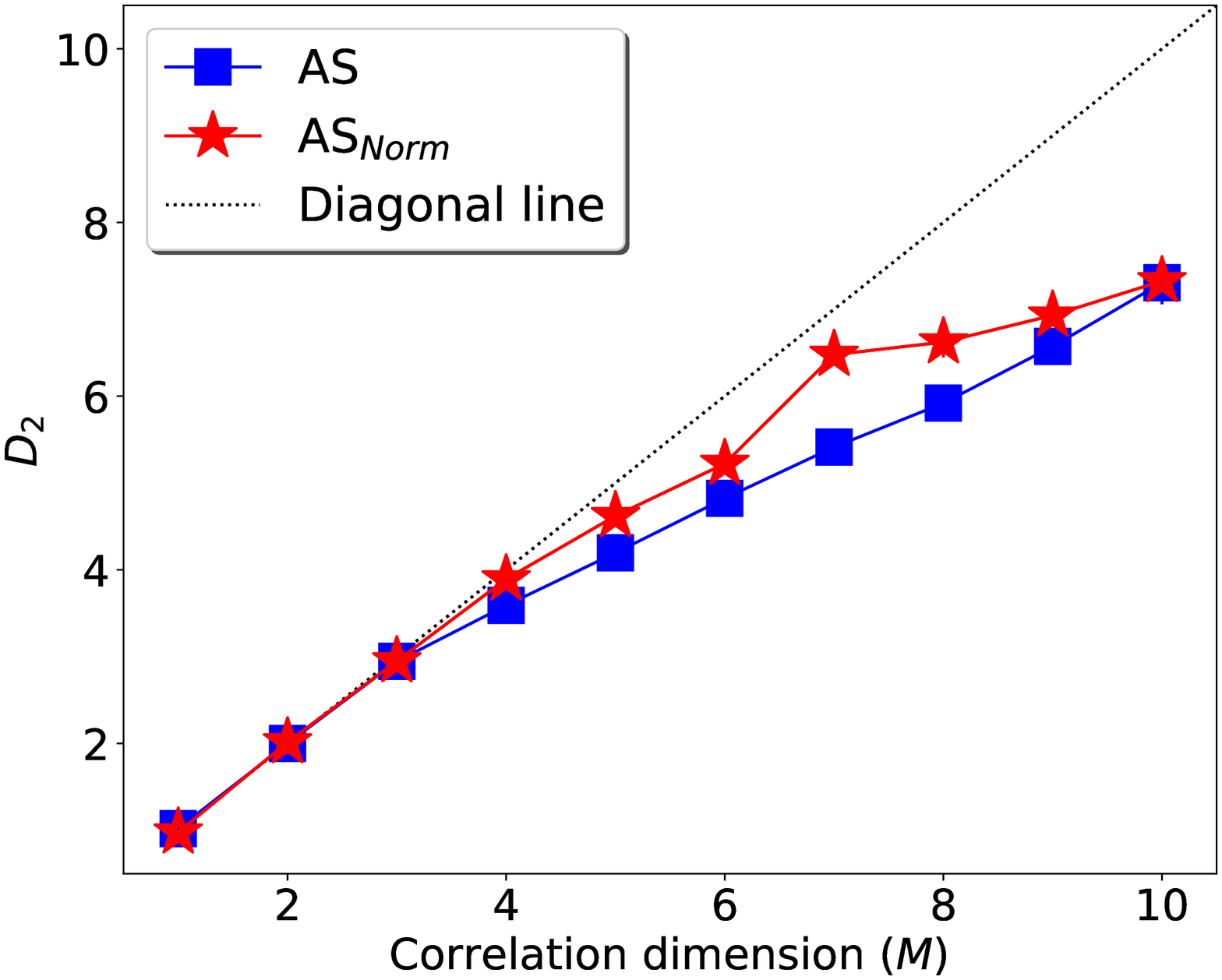}
\caption{
Variation of correlation dimension ($D_2$) with the embedding dimension ($M$) for $AS$ (square points) and $AS_{\rm Norm}$ (asterisks) calculated from hemispheric sunspot numbers.}
\label{fig:MvsD2_HSSN}
\end{figure}

The sunspot number is probably the longest data available to study solar activity. However, the hemispheric data of sunspot number
were available only from 1992. 
Recently, hemispheric sunspot number data has been derived and made publicly available 
\citep{Veronig21}\footnote{https://wwwbis.sidc.be/silso/datafiles} during 1874--2020. As the sunspot number has 
been extensively used to study solar activity and it 
has a strong correlation with the sunspot area number, 
we present the results of our time-series analyses for the 
$AS$ and $AS_{Norm}$ computed from the sunspot number during the period May 1874 - September 2016. 
The results are presented in \Tab{tableB1} and \Fig{fig:MvsD2_HSSN}.

\begin{table}
\centering
\caption{
Summary of results obtained for Higuchi's dimension and Hurst exponents from hemispheric sunspot numbers.} 


\begin{tabular}{llcc}
\hline \hline
\multicolumn{1}{l}{Method} & \multicolumn{1}{l}{Window} & \multicolumn{2}{c}{Time Series}                                                                        \\ \cline{3-4} 
                           & (months)                                                                      & AS                   & AS$_{Norm}$         \\ \hline
        Higuchi                   &          2-55                                                                     & 1.90                                     & 1.90 \\ \hline
\multicolumn{1}{c}{Hurst (R$/$S)}     &11-110                                                    &    0.83                                          &   0.83 \\
                           & 263-650                                                                       & 0.93 & 0.83 \\ \hline
\multicolumn{1}{c}{Hurst (DFA)}    & 11-110                                                    &         0.90                                                  &       0.89          \\
                           & 263-650                                                                       & 1.07                                     & 0.88 \\ \hline
\multicolumn{1}{c}{Hurst (PR)}     & K = 28                                                    &    0.73                                &      0.76            \\ \hline
\end{tabular}
\label{tableB1}

\end{table}

\bsp    
\label{lastpage}
\end{document}